\documentclass[11pt]{article}
\usepackage{amsfonts,amssymb}
\usepackage{mathptmx}
   \newcommand{\A}{{\tilde A}}
     \newcommand{\F}{{\tilde F}}
\renewcommand{\a}{{\alpha}}
\newcommand{\C}{{\bf C}}
\newcommand{\Z}{{\bf Z}}
\newcommand{\ra}{{\rightarrow}}
\newcommand{\la}{{\leftarrow}}
\newcommand{\qed}{{\,\,\,\em{Q.E.D.}}}

\newtheorem{thm}{Theorem}
\newtheorem{prop}{Proposition}
\newtheorem{lmm}{Lemma}

\font\stepthree=cmb10 scaled\magstep4
\def\ce{\centerline}

\def\b{\bigskip}
\def\bb{\bigskip\bigskip}

\def\no{\noindent}

 \relax

\def\~{\hskip-1mm}

\def\harr#1#2{\smash{\mathop{\hbox to .25 in{\rightarrowfill}}
    \limits^{\scriptstyle#1}_{\scriptstyle#2}}}

\font\aab=cmbsy10 at 18pt
\def\bigbullet{\lower.3ex\hbox{{\aab  \char'017}}}

\def\A{\hbox{$A(\langle q\rangle,a)$}}

\begin{document}

\title{Quantization on Curves}

 \author  {Christian Fr\o nsdal\\
{\it Physics and Astronomy Department, University of California,}\\
{\it Los Angeles, CA 90095-1547, USA,} 
\texttt  {fronsdal@physics.ucla.edu}\\
{\small With an appendix by}\\ {Maxim Kontsevich}\\
{\it Institut de Hautes Etudes Scientifiques,}\\
{\it 35 route de Chartres, F-91440 Bures sur Yvette, France,}
\texttt {maxim@ihes.fr}}
  
\date{\small Received 30 June 2002, appendix added 7 June 2004, 
revised 12 September 2006}
\maketitle

\begin{abstract}
 Deformation quantization on varieties with
singularities offers perspectives that are not found on manifolds.
The Harrison component of Hochschild cohomology, vanishing on smooth
manifolds, reflects information about singularities. The Harrison
2--cochains are symmetric and are interpreted in terms of
abelian $*$--products. This paper begins a study of abelian
quantization on plane curves over $\mathbb{C}$,
being algebraic varieties of the form ${\mathbb{C}}^2/R$, where $R$ is a
polynomial in two variables; that is, abelian deformations of the
coordinate algebra $\mathbb{C}[x,y]/(R$).
To understand the connection between the singularities of a variety and
cohomology we determine the 
algebraic Hochschild (co-)homology and its Barr--Gerstenhaber--Schack
decomposition. Homology is the
same for all plane curves $\mathbb{C}[x,y]/R$, but the
cohomology depends on the local algebra of the singularity of $R$ at the
origin.
\end{abstract}



   \section {Introduction.}
Deformation quantization is a term coined by Moshe Flato, who suggested
that any nontrivial associative deformation of an algebra of functions
should be interpreted as a kind of ``quantization". Deformation
quantization is  \cite{BFFLS} the study of associative $*$--products 
of the form $f*g = fg + \sum_{n>0} \hbar^nC_n(f,g)$,
where $\hbar$ is a formal  parameter. This concept has gained wide
currency and has  been intensively developed in recent years, but almost
exclusively in the context of smooth Poisson manifolds \cite{Fe,Kon,Kon2}. In
that case it is natural to consider deformations ``in the direction of the
Poisson bracket" (Drinfel'd); that is, taking $C_1(f,g) = \{f,g\}$,
which is of course antisymmetric. But even if more general deformations
were to be considered, independent of the symplectic structure,
antisymmetry of $C_1$ entails no essential loss of generality for
quantization on  a smooth (finite dimensional) manifold. A famous result of
Hochschild, Kostant and Rosenberg \cite{HKR} implies that any $*$--product 
on a regular, commutative algebra is equivalent to one with 
antisymmetric $C_1$. For a related `smooth' result, see \cite{V}.

It would seem, therefore, that the time has come to study
deformation quantization on varieties with singularities. The
cohomological implication of singularities should be interesting.

The Hochschild complex of any commutative algebra   decomposes into
smaller complexes; in the case of an algebra $A$ generated by $N$
generators, into $N$ subcomplexes \cite{B,Fl,GS}. The topology of a smooth
manifold is  related to the restriction of the Hochschild complex to
alternating maps $A^\wedge \rightarrow A$, dual to simplicial homology,
and the only component with non-vanishing cohomology. But on varieties
with singularities other components of the Hochschild complex come into
play, which suggests the use of cohomological methods for the study of
singularities.

Examples of quantization on singular varieties had been known in
connection with geometric quantization (and $*$--quantization) on
coadjoint orbits of Lie algebras, but the cohomological implications had
not been recognized. (See \cite{BFFLS,Fr,Fr2}.) The connection between 
singularities and cohomology was studied by Harrison \cite{H}, who was 
the first to describe the component of Hochschild cohomology that has 
become known, if not widely known, as Harrison cohomology. The 2--cochains 
of this complex are symmetric. On a commutative algebra every exact Hochschild
2--cochain is symmetric, so that triviality
is not an issue if $C_1$ is antisymmetric. But it is an important
consideration in the case of abelian $*$--products.
\b

\ce {\bf The BGS idempotents.}

The $p$--chains of the Hochschild homology complex of a commutative algebra $A$
are the $p$--tuples $a = \sum a_1\otimes \cdots \otimes a_p \in A^{\otimes p}$, 
and the differential is defined by
$$
da = a_1a_2\otimes a_3 \otimes \cdots \otimes a_p - a_1\otimes
a_2a_3\otimes a_4\cdots \otimes a_p + \cdots +(-)^pa_1\otimes \cdots
a_{p-2}\otimes a_{p-1}a_p.
$$
The $p$--cochains are maps $A^{\otimes p}   \rightarrow  A$,  and the
differential is
$$
\delta C(a_1,\cdots ,a_{p+1}) = a_1C(a_2,\cdots ,a_{p-1}) - C(da) -
(-)^pC(a_1,\cdots ,a_p)a_{p+1}.
$$
     After the pioneering work of Harrison \cite{H} and Barr \cite{B}, 
the complete decomposition of the Hochschild cohomology of a
commutative algebra was found by Gerstenhaber and Schack \cite{GS}. The
Hochschild cochain complex splits into an infinite sum of direct summands.
(If the algebra is generated by $N$ generators then there are only
$N$ nonzero summands.)  The decomposition is based on the action of $S_n$
on $n$--cochains, and on the existence of  $n$ idempotents
$e_n(k),~k = 1,\cdots n,$~in $\mathbb{C} S_n$,   $\sum_ke_n(k) = 1$,  with
the property that
$
\delta\circ e_n(k) = e_{n+1}(k)\circ \delta.
$
Thus we have Hoch$_n = \sum_{k = 1}^nH_{n,k}$, ~Hoch$^n = \sum_{k =
1}^nH^{n,k}$ with $H_{n,1} = $Harr$_n$ and $H^{n,1} =$  Harr$^n$.

A  generating function was found by Garsia \cite{Ga},
$$
\sum_{k = 1}^nx^ke_n(k) = {1\over n!}\sum_{\sigma\in S_n}
(x-d_\sigma)  (x-d_\sigma +1)  \cdots   (x-d_\sigma +n-1){\rm sgn}(
\sigma)\sigma,
$$
where $d_\sigma $ is the number of descents,~~$\sigma(i) > \sigma(i+1)$,
~~in $\sigma (1 \cdots n)$. \footnote {Example: $\sigma(1234) = 3142$ 
has one descent, from 2 to 3.} The simplest idempotents are
\begin{eqnarray*}
e_2(1) 12  &=& {1\over 2}(12 + 21),\\
e_3(1) 123&=& {1\over 6}\bigl(2(123 - 321)  + 132 - 231 + 213 - 312
\bigr), \\ e_3(2) 123&=& {1\over 2}(123 + 321)\\
e_n(n)& =& {1\over n!}\sum_{\sigma\in S_n}{\rm sgn}( \sigma)\sigma.
\end{eqnarray*}
The Hochschild chains decompose in the same way, with $d\circ e_n(k) =
e_{n-1}(k)\circ d$.

\b

\ce {\bf Summary.}

Section 2 is concerned with abelian $*$--products on an arbitrary plane
curve. The space of equivalence classes of first order abelian deformations 
of the algebra of polynomials on $\mathbb{C}[x,y]/(R)$ is isomorphic to 
the local algebra of the singularity of $R$ at $x = y = 0$. The
Harrison component Harr$^3 = H^{3,1}$ of Hoch$^3$ vanishes, which implies
that there are no obstructions to continuing a first order 
abelian $*$--product to higher orders.
In this paper the strategy that leads to the calculation of Hochschild
cohomology calls for a preparatory investigation of a homological complex
that is not strictly Hochschild, but rather its restriction 
$A \rightarrow A_+$ to the non-unital subalgebra
$A_+$ of  positive degree; this has no effect on the cohomology.

   In Section 3 the  Hochschild homology is calculated for the case of a
plane curve, with its BGS decomposition.
In Section 4 the Hochschild cohomology is investigated; the result in
Theorem 4.9. Section 5 contains a detailed calculation of the BGS 
decomposition for the singularity of $x^n = 0$ at $x = 0$.

The Appendix, by Maxim Kontsevich, explains in modern mathematical
language a way to calculate Hochschild and Harrison cohomology groups for 
algebras of functions on singular planar curves etc. based on Koszul 
resolutions. 

\section{ Associative  $*$--products and cohomology.}

\no{\it 2.1 Formal   $*$--products.} A formal, abelian
$*$--product on a commutative algebra $A$   is a commutative,
associative product on the space of formal power series in a formal parameter
$\hbar$  with coefficients in $A$, given by a formal series
$$
f*g = fg + \sum_{n>0} \hbar^nC_n(f,g).\eqno(2.1)
$$
Associativity is the condition that $f*(g*h) = (f*g)*h $, or
$$
\sum_{m,n = 0}^k \hbar^{m+n}\biggl(
C_m(f,C_n(g,h))-C_m(C_n(f,g),h))\biggr) = 0,\eqno(2.2)
$$
where $C_0(f,g) = fg$. This must be interpreted as an identity in
$\hbar$; thus
$$
\sum_{m,n = 0}^k \delta_{m+n,k}\biggl(
C_m(f,C_n(g,h))-C_m(C_n(f,g),h) )\biggr) = 0, ~~k = 1,2,\cdots~.\eqno(2.3)
$$
The  formal $*$--product (2.1) is associative to order $p$ if Eq(2.3)
holds for $k = 1,\cdots p$.

A first order abelian $*$--product is a product
$$
f*g = fg + \hbar C_1(f,g),\quad C_1(f,g) = C_1(g,f),\eqno(2.4)
$$
associative to first order in $\hbar$, which is the requirement that
$C_1$ be closed,
$$
\delta C_1(f,g,h) := fC_1(g,h) - C_1(fg,h) + C_1(f,gh) - C_1(f,g)h = 0.
$$

Suppose that a formal $*$--product is associative to order $p \geq 1$;
this statement involves $C_1,\cdots ,C_p$ only, and we suppose these
cochains fixed. Then the condition that must be satisfied by
$C_{p+1}$, in order that the $*$--product be associative to order $p+1$, is
$$
\sum_{m,n = 1\atop m+n = p+1}^p  \biggl(
C_m(f,C_n(g,h))-C_m(C_n(f,g),h) )\biggr) = -\delta C_{p+1}(f,g,h).\eqno(2.5)
$$
The left hand side is closed, and thus it is seen that the obstructions to
promote associativity from order $p$ to order $p+1$ are in Hoch$^3$.

There is an important difference between the
two cases of symmetric and antisymmetric $C_1$. If $C_1,\cdots ,C_p$ are
symmetric, then the left hand side of (2.5) has the symmetry of the
idempotent $e_3(1)$ (a Harrison cochain) and it is the symmetric part of
$C_{p+1}$ that is relevant, while the antisymmetric part of $C_{p+1}$
must simply be closed.  Symmetry of the $*$--product can therefore be
maintained to all orders. If   $C_1$ is antisymmetric, and $p = 1$, then
the left hand side has the symmetry of $e_3(1) + e_3(3)$. The first part
must be balanced on the right hand side by means of the symmetric part of
$C_2$; the second part must vanish, and this condition is the Jacobi
identity for $C_1$.

The obstructions against continuing a formal, first order,  
abelian $*$--product to higher orders are in Hoch$^3 $; 
more precisely, they are in $H^{3,1} =$ Harr$^3(A,A)$.

A formal $*$--product is trivial if there is an invertible map
$E:A\rightarrow A$, in the form of a formal series
$E(f) = f + \sum_{n>0}\hbar^nE_n(f)$ such that
$E(f*g) =  E(f)E(g)$.
A first order, abelian $*$--product is trivial if there is a 1--cochain
$E_1$ such that
$$
C_1(f,g) = \delta E_1(f,g) = fE_1(g) - E_1(fg) + E_1(f)g.
$$

   \b
\no{\it 2.2. Deformations on a curve.} In view of the
theorem of Hochschild, Kostant and Rosenberg \cite{HKR} cited earlier, there
can be no nontrivial, abelian  $*$--products on a smooth manifold. It is
natural to turn to varieties with singularities, and especially algebraic
varieties. It is the aim of this
paper to explore the phenomena, with elementary methods of calculation,
in the case of plane curves over $\mathbb{C}$,
$M = \mathbb{C}^2/R$, where $R$ is a $\mathbb{C}$--polynomial. 
The algebras of interest are the coordinate algebra
$$
A = \mathbb{C}[x,y]/(R),\eqno(2.6)
$$
with generators $x,y$ and a single polynomial relation $R$. The polynomial 
$R$ can be transformed, by a linear change of variables, to either of the
forms
$R = x^m -  P(x,y)$ or $R = y^n -Q(x,y)$,
where the polynomial $ P$ is of order less than $m$ in $x$ and the
polynomial $Q$ is of order less than $n$ in $y$. Either form gives rise to
a Poincar\'e--Witt basis for $A$, for example,
$x^iy^j$, $i = 0,1,\cdots \infty$, $j = 0,1,\cdots ,n-1$.

The deformed algebra has a Poincar\'e--Witt basis of the same form. 
Let $W$ be the map that takes a $*$--monomial of
this basis to the same ordinary monomial of the original basis. Let
$R_\hbar := W(R^*)$ and let $M_\hbar :=\mathbb{C}^2/R_\hbar$.
Then, morally, the $*$--product is trivial if there is a bijection
$E:  M_\hbar \rightarrow M$ such that $R_\hbar \mapsto R$.
However, since $\hbar$ is a formal parameter, the following definition is
preferred.
\b

\no{\it 2.3. Definition.  A $*$--product, as defined in this section, is
trivial if there is a mapping  by a formal power series in 
$\hbar,~ E = {\rm Id} + \sum_{n>1}\hbar^nE_n$, such that $R_\hbar \mapsto R$.}
\b

\no{\it 2.4. First order $*$--product on a curve.} Consider a first order,
associative and abelian $*$--product on the algebra (2.6), 
with the polynomial $R$ in the form $R = y^n - Q(x,y)$. 
A change of variables ensures that $(x^iy^j)*(x^ky^l) = x^{i+k}y^{j+l}$  
for $j+l<n$  and
$$
y^i*y^{n-i}   =  Q(x,y)  + \hbar   Q_1(x,y),\quad 1\leq i\leq n-1,\eqno(2.7)
$$

The first order deformation (2.7) is trivial if there is a derivation $E$
such that $Q_1 = E(R)$. See Subsection 4.6.
   \b

\no{\it 2.5. Example.} Let $A = \mathbb{R}[x,y]/(R), ~ R =
y^2-x^2 - r^2,~~ r^2 \in \mathbb{C}$, decompose $f\in A$ as
$f = f_+ + yf_-$, $f_\pm\in \mathbb{R}[x]$,
and define a $*$--product on $A$ by setting $f*g = fg + \hbar f_-g_-$.
Then $Q_1 = 1$ and we seek $E$ such that $E(x^2 + r^2 -
y^2) = 1$. The general solution to this equation is
$2E = {-1\over r^2}(x\partial_x + y\partial_y) +
\alpha(y\partial_x +x\partial_y)$, with $\alpha\in A$.

Of course, this breaks down if $r^2 = 0$, and the simple reason 
why there is no solution in this case is that there is no differential 
operator $E$ such that the polynomial $E(x^2-y^2)$ contains a constant term.
\b

\no{\bf 2.6. Proposition.} {\it Let $X$ be the space of polynomials in $x$
and $y$, of degree less than $n$ in $y$, and let $DR$ be the gradient ideal 
of $R$. As vector spaces, $X$ coincides with $A$ and $DR$ consists of all
differentials of $R$. The space of equivalence classes of essential, 
first order $*$--products on $A$ is the space $X/DR$}, \
Harr$^2(A,A) = X/DR$.
\b

\no{\it 2.7. Example.} Let $M = \mathbb{C}^2,~~R = y^2-x^3$. A full set of
representatives of $X/DR$ is $ a+ bx, ~a,b \in \mathbb{C}$. The deformed
algebras are $A_\hbar = \mathbb{C}[x,y]/R_\hbar$ with
$R_\hbar =  y^2-x^3-\hbar(ax+b)$.
Expand $f(x,y) = f_+(x) + yf_-(x)$. Then
$f*g = fg + \hbar C_1(f,g)$, where $C_1(f,g) = (ax+b)f_-g_-$.

\section{Homology.}

This section deals with the homology of a modified Hochschild complex.
The strategy that is used in this paper, to calculate the Hochschild
cohomology of $A$, begins by a determination of the homology of the
algebra $A_+$, the subalgebra with positive degree of $A$.
    The $n$--chains of  this  homology of $A_+$ are $n$--tuples
$a = a_1\otimes a_2\otimes \cdots a_n,~~ a_i \in A_+,~~i = 1,\cdots N$.
\b

\no{\it 3.1. 2--chains.} Every `Hochschild' 2--chain is  homologous to a
2--chain of the form $x\otimes a + y \otimes b$.
It will be convenient to re-label the generators, $x,y \mapsto x_1,x_2$, then
$a \approx \sum x_i\otimes a_i$, 
$a_i \in A_+,~~i = 1,2~$.
It is closed if $\sum x_ia_i =  0$. We shall suppose that $R$ has 
no constant term and no linear terms, then $a$ has the representation
$$
a  \approx \sum x_i \otimes x_j\epsilon^{ij}b + \sum_{i=1}^2 x_i\otimes R_i c,
$$
where $\epsilon^{ij} = -\epsilon^{ji}, \epsilon^{12} = 1,~\sum x_iR_i = R$
and where $b,c$ are in the unital augmentation $A$ of $A_+$. The first
term  is exact if $b\in A_+$, the second term is exact if $c \in A_+$
and (a section of) $H_2 = Z_2/B_2$ is spanned (over $\mathbb{C}$) by the chains
$
x_1\wedge x_2  ~~{\rm and } ~~\sum x_i \otimes R_i .
$
The second one is homologous to a symmetric chain that is a
basis for Harr$_2 = H_{2,1}$.
\b

\no{\it 3.2. Example.} If $R = y^2 - x^n$, then ${\rm Harr}_2$ has
dimension 1 and every symmetric, closed 2--chain is homologous to a
$\mathbb{C}$--multiple of $x\otimes  x^{n-1} + x^{n-1}\otimes x-2y\otimes y$.
\b

\no{\it 3.3. 3--chains.} Every 3--chain is homologous to one of the form
$a = \sum x_i\otimes b_j\otimes c^{ij}$. If $a$ is closed it takes the form
$a \approx  \sum x_i \otimes x_j\epsilon^{ij}b\otimes b' +
x_i\otimes R_i c\otimes c',~~b,c \in A$
which is homologous to
$a \approx  \sum x_i \otimes x_j\epsilon^{ij} \otimes bb' +
x_i\otimes R_i  \otimes cc'$, with
$x_2bb' + R_1cc' = 0$ and $-x_1bb' + R_2cc' = 0$.
A simple case-by-case study shows that we then have:
$$
bb' = \alpha R_1 + \beta R_2,\quad cc' = -\alpha x_2 + \beta x_1,
$$ 
with $\alpha, \beta$ in  $A$. Thus any closed 3--chain is homologous 
to one of the form
$$
\biggl( (x_1 \wedge x_2) \otimes R_1c_1 -\sum x_i\otimes 
R_i\otimes x_2c_1\biggr)  - \biggl((x_1\wedge x_2)\otimes R_2c_2 + 
\sum x_i\otimes R_i\otimes x_1c_2\biggr).\eqno(3.1)
$$
The first (second) term is exact unless $c_1 (c_2)$ is in $\mathbb{C}$.
Adding an exact, alternating 3--cycle we get an alternative section 
of $Z_3/B_3$ with a basis that consists of the two chains
(the GS idempotents were defined in the introduction)
\begin{eqnarray*}
\alpha_1 &=&   e_3(2) \bigl( x_1\otimes x_2\otimes R_1 -
x_2\otimes R_1\otimes x_1 - x_2\otimes x_1\otimes R_1 - x_2\otimes
R_2\otimes x_2\bigr),  \\
   \alpha_2 &=&   e_3(2) \bigl( x_2\otimes x_1\otimes R_2-
x_1\otimes R_2\otimes x_2 - x_1\otimes x_2\otimes R_2 - x_1\otimes
R_1\otimes x_1\bigr).\quad (3.2)
\end{eqnarray*}
   Thus Hoch$_3 = H_{3,2}$ has dimension 2 and Harr$_3= 0$.
\b

Another way to reach this conclusion is to differentiate (3.1). The
result is\break
$(c_1x_2 + c_2x_1)\wedge R$, which is in $ Z_{2,2}$ and which
implies that (3.1) $\in~ Z_{3,2}$.
\b

\no{\it 3.4. Example.} If $R = y^2 - x^2$, set $u = x+y,~v = x-y$. The
basis (3.2) is then $\{u\otimes v\} \otimes u$, $v\otimes \{u\otimes v\}$ 
and the dimension of Hoch$_3$ is 2.  
More precisely, dim~$H_{3,k}$ is $0,2,0$ for $k=1,2,3$.
\b

\no {\it 3.5. Example.} If $R = y^2 - x^3$, then the chains (3.2) become
$$
y\otimes x\otimes y - x\otimes y\otimes y - y\otimes y \otimes x +
x\otimes x^2 \otimes x
$$
and
$$
e_3(2)\bigl( x\otimes y\otimes x^2 - y\otimes x^2\otimes x - x^2\otimes
x\otimes y + y\otimes y \otimes y\bigr).
$$

It is straightforward to prove the following.
\b

\no{\bf 3.6.  Proposition.} {\it Let 
$P^1  = \{x_1,x_2\}$,  
$P^{n+1} = P^n \otimes M_n$, and
$$
M_{2k+1}  =  \pmatrix{R_1&-x_2\cr  R_2 & x_1},\quad
M_{2k}  = \pmatrix{x_1 & x_2\cr -R_2&R_1}.
$$
Then for $n>1$ every closed $n$--chain is homologous to an $n$--chain in
the linear span of the two linearly independent polynomials in $P^n$.}
\b

\no{\it 3.7. Example.} If $R = y^2 - x^2$, set $u = x+y,~ v = x-y$. The
dimension of Hoch$_n$ is 2; the basis is
$\{u\otimes v\otimes   u \cdots ,~~ v\otimes u\otimes v\otimes u\cdots\}$.
\b

\no{\bf 3.8. Theorem.} {\it Hoch}$_{2k} = H_{2k,k} + H_{2k,k+1}$, {\it each
component one-dimensional~over $\mathbb{C}$}, and
{\it Hoch}$_{2k-1} = H_{2k-1 ,k}$, {\it two-dimensional over 
$\mathbb{C}$}, $k = 1,2,...$~.
\b

\no{\it Proof.} For $k = 1,...,p-1$, 
$P^{p+1} = P^k\otimes M_k\otimes M_{k+1}\otimes ... \otimes M_p $
and thus
$$
dP^{p+1} = P^1M_1\otimes M_2\otimes ... \otimes M_p +
\sum_{k=1}^{p-1}(-)^kP^k\otimes M_kM_{k+1} \otimes
...\otimes M_p.
$$
We have $M_kM_{k+1} = R$ times the unit matrix and $P^1M_1\otimes M_2 =
R\otimes P^1$;
consequently  $dP^1 = 0,~ dP^2 = \{R,0\}$ and
$dP^{p+1} = R \bigcirc \hskip-4mm {\scriptstyle sh}~ P^{p-1},~~ p\geq 2$.
If $a \in  C_{p,k}$, then $da\in  C_{p-1,k}$, and
$R\bigcirc \hskip-4mm {\scriptstyle sh}~ a$ is homologous to some $b \in
C_{p+1,k+1}$. Hence if $P^{p-1}
\in C_{p-1,k}$, then
$P^{p+1}$ is homologous to a $C_{p+1,k+1}$ chain.The action of these maps
between spaces with cohomology
is shown in the diagram.
$$
\matrix{&   C_{2,1} \cr  &\hskip-8mm\swarrow  &  \hskip-10mm\searrow  \cr
C_{1,1}  && C_{3,2}\cr
&\hskip-8mm\searrow&\hskip-10mm\swarrow&\cr & C_{2,2} \cr}\cdots~~
\matrix{&   C_{2k,k} \cr  &\hskip-15mm\swarrow  &  \hskip-15mm\searrow  \cr
C_{2k-1,k}  &&
C_{2k+1,k+1}\cr &\hskip-15mm\searrow&\hskip-15mm\swarrow&\cr & C_{2k,k+1}
\cr}\eqno(3.4)
$$
A southeast arrow represents the map $a \mapsto  R\bigcirc \hskip-4mm
{\scriptstyle sh}~ a$; a southwest
arrow is the action of the differential.
The projections of $\{P_i^{2k+1}\}_{i = 1,2}$ form a basis for
$H_{2k+1,k+1}$ and the projections of
$P_1^{2k}$ (resp. $P_2^{2k}$) are bases for $H_{2k,k}$ (resp. $H_{2k,k+1}$).
\bb

\section{Cohomology.}

\no{\it 4.1. The reduction process.} The chains considered in this
section are restricted to positive degree. The cochains are valued in
$A$.   A p--cochain is closed if
$$
\delta C(a_1,\cdots ,a_{p+1}) = a_1C(a_2,\cdots ,a_{p+1}) - C(da)
-(-)^pC(a_1\cdots ,a_p)a_{p+1} = 0.\eqno(4.1)
$$
One may attempt to interpret this relation as fixing the value $C(da)$,
recursively in the degree of the argument.  The obstruction to this  is
$da = 0$, but if $a$ is exact then (4.1) is satisfied automatically by
virtue of its being true for arguments of lower degree. (One can show that,
in this context, if $a$ is exact then there is $b$ of the same degree
such that $a=db$.) It is enough, therefore, to verify closure for a basis 
of representatives of Hoch$_{p+1}$.

A closed $p$--cochain $C$ is a coboundary if there is a $(p-1)$--cochain $E$ 
such that
$$
C(a) = a_1E(a_2,\cdots , a_p) - E(da) + (-)^pE(a_1,\cdots
a_{p-1})a_p.\eqno(4.2)
$$
This relation can be solved for $E(da)$, recursively by
increasing degree, except for the obstruction presented by $da = 0$.
But if $a = db$ then  $C(a)$ is determined by $\delta C(b) = 0$. So it is
enough to examine (4.2) for a complete set of representatives of Hoch$_p$.

The most useful interpretation is this. Given any closed $p$--cochain
a ``gauge transformation" is the addition of an exact $p$--cochain,
$C\rightarrow C + \Delta C$, with
$$
\Delta C(a_1...a_p) = a_1E(a_-) +(-)^pa_pE(a_+) - E(da).\eqno(4.3)
$$
The space Hoch$^p$ is the spac of closed, gauge-invariant
$p$--cochains.
\b

{\it If any BGS component $H_{p,k}$ of Hoch$_p$ vanishes 
then the corresponding component $H^{p,k}$ of Hoch$^p$ is zero. 
There are no obstructions to continuing a first order, 
abelian $*$--product to higher orders.}
\b

   \no{\it 4.2. Closure for $p = 1$.} 
The 2--homology is spanned by $x_1\wedge x_2$ and $x_i\otimes R_i$. 
We shall replace the latter by
$ \hat R = \sum A_{ij}x_1^i\otimes x_2^j,~~ R = \sum A_{ij}x_1^ix_2^j$. 
The relation $\delta C(x_1\wedge x_2) = 0$ is trivial. The formula
$\delta C(x_1^i\otimes x_2^j) = x_1^iC(x_2^j)+x_2^jC(x_1^i) - C(x_1^ix_2^j)$ 
tells us that, if $C$ is closed, then for any polynomial $f$, 
$C(f) = C(x_i)\partial_if$. Hence (this is the result 2.6)
$$
\delta C(P_1^2) = C(x_i)\partial_i R,~~\delta C(P_2^2) = 0.\eqno(4.4)
$$

For the algebra $\mathbb{C}[x,y]$, $Z^1$ is the space of vector fields with
coefficients in the unital augmentation of the same algebra, 
but for $A = \mathbb{C}[x,y]/R$, $Z^1$ is the algebra of vector fields
that annihilate $R$ (the algebra of vector fields tangential to the curve).
\b

\no{\it 4.3. Closure for $p = 2$.} For homology we use the basis (3.3); it
is enough to examine one of the two,
$$
P_1^3  = \hat R \otimes x_1 + x_1\wedge x_2\otimes R_2,
$$
$$
\delta C(P^3_1)  = x_1C(R_1\wedge x_1) + x_2C(R_2\wedge x_1) -
R_2C(x_1\wedge x_2).
$$
The first two arguments are exact; a certain amount of calculation is
needed to verify that these terms are of the same form as the third one. 
We need the following simple formula, satisfied by closed 2--cochains:
$C(x_2\wedge f)  = C(x_2\wedge x_1)\partial_1f$, $f \in A$.
Now it follows easily that
$\delta C(P^3_1) = -C(x_1\wedge x_2) \partial_2 R$, $\delta C(P^3_2) =
C(x_1\wedge x_2) \partial_1 R$. 
Therefore, we can interpret the condition $\delta C(a) = 0$ as fixing
the value $C(da)$, provided only that
$C(P_2^2)\partial_iR = 0$, $i = 1,2$. 
(That is satisfied if $R= x^2y^3$, $C(x\wedge y) = xy$.)
 \b

\no{\bf 4.4.  Theorem.} {\it Closure of a $p$--cochain $C$ implies that its
values for exact arguments are  given recursively in the polynomial degree 
as in (4.1). Conversely, (4.1) can be solved
recursively for all $C(da)$, if and only if the following conditions hold}
\begin{eqnarray*}
C\in Z^{2k,k+1}&:&~~~~~C(P_2^{2k})\partial_iR = 0,~ i = 1,2;\cr
C \in Z^{2k+1,k+1}&:&~~~~\sum C(P_i^{2k+1})\partial_iR =0;\cr
C\in Z^{2k,k}&:&~ ~~~~~~~~~~~~~~~~~~~{\it always}.\end{eqnarray*}

\no{\it 4.5. Gauge invariance for $p = 1$.} Trivial, all 1--cochains are
gauge invariant, $H^1 = Z^1$.
\b

\no{\it 4.6. Gauge invariance for $p = 2$.} We must examine evaluations on
the homology basis. To begin with, $\Delta C(x_1\wedge x_2) = 0$,
so that the evaluation $C(x_1\wedge x_2)$ is gauge invariant.  To examine
the supplementary homology space, set $R = \sum A_{ij}x_1^ix_2^j$, 
$\hat R = \sum A_{ij}x_1^i\otimes x_2^j$. Then we have
$$
\sum  {\scriptstyle 1\over \scriptstyle 2}A_{ij}\bigl(\Delta
C(x_1^i\otimes x_2^j) + x_1^i\sum_{k =
2}^{j-2}x_2^k\Delta C(x_2 \otimes x_2^{j-1-k}) + x_2^j\sum
_{k=0}^{i-2}\Delta C(x_1\otimes
x_1^{i-k-1})\bigr)
$$
$$
= E(x_i)\partial_iR.
$$

Hence, in a gauge where $C$ vanishes on arguments of lower degrees,
$\Delta C(\hat R) \in DR$ and we have recovered Proposition 2.6.
\b

\no{\it 4.7. Gauge invariance for $p = 3$.} We have
\begin{eqnarray*}
\delta C(P_1^3) &=& \Delta C(\hat R \otimes x_1 + x_1\wedge x_2\otimes R_2) \\ 
& =& x_1E(R_1\wedge x_1 ) + x_2E(R_2\wedge x_1)) - R_2E(x_1\wedge x_2)
\qquad\qquad\qquad \qquad(4.5) \\
&=& \sum {\scriptstyle 1\over \scriptstyle
2}A_{ij}\bigl\{x_1E(x_1^{i-1}x_2^j\wedge x_1) +
x_2E(x_1^ix_2^{j-1}\wedge x_1)\bigr\}- R_2E(x_1\wedge
x_2). \end{eqnarray*}

\no With the help of the identity
$$
\sum_{k = 1}^{i-1} x_1^k\Delta C(x_1\otimes x_1^{i-k-1}x_2^j \otimes x_1) =
x_1^iE(x_2^j\wedge x_1) - x_1E(x_1^{i-1}x_2^j\wedge x_1), ~ j \geq 1,
$$
and another one, similar, we can reduce (4.5) to
$$
\Delta C(P^3_1)    + \sum_{k = 1}^{i-1}A_{ij} x_1^k\Delta C(x_1
\otimes x_1^{i-k-1}x_2^j \otimes x_1)
+\sum_{k = 1}^{i} A_{ij}x_2x_1^{k-1}\Delta C(x_1\otimes x_1^{i-k}x_2^{j-1}
\otimes x_1)
$$
$$
   = \sum A_{ij}\bigl\{x_1^iE(x_2^j\wedge x_1) + x_1^iE(x_2^{j-1}\wedge
x_1)\bigr\} - R_2 E(x_1\wedge x_2).
$$
A similar, further  reduction leads to the result that, if $\delta C$
vanishes on arguments of lower orders, 
$\Delta C(P_1^3) + ... = -(\partial_2R)E(x_1\wedge x_2)$,
$\Delta C(P_2^3) + ... = (\partial_1R)E(x_1\wedge x_2)$.
We recall that $\Delta C(a) = \delta E(a_1)$ and remember from Subsection~4.3 
that $\delta E = 0$ implies that $ \partial_i RE(x_1\wedge x_2) = 0$. 
The above result is thus natural; the calculation is needed only to fix 
the numerical coefficients.
\b

\no{\it 4.8. Proposition. If the gauge is fixed by the condition that $
C(a) = 0$ for arguments $a$
of lower degree, then the remaining gauge transformations take the
following form,}
\begin{eqnarray*}
\Delta C(P^1) &=& 0, \quad \Delta C (P_1^{2k}) = \sum E_i\partial_iR,\\
\Delta C(P_2^{2k}) &=& 0, \quad \Delta C(P^{2k+1}) = EdR^*,~~ k > 0.
\end{eqnarray*}

\no{\it Proof (outline).} (a) The statement reflects the structure of
(3.4). The dimension of $H^{p,k}$, over the local algebra, more or less, 
coincides with the dimension of $H_{p,k}$. `More or less' comes
from the existence of homologies of lower orders, as the complete
calculation in Subsection 4.7 shows.

(b) We have
\begin{eqnarray*}
\Delta C(P_1^{2k}) &=& \sum R_iE(P_i^{2k-1}) + \sum x_iE(Q_i) + ...~, \cr
\Delta C(P_i^{2k+1}) &=& \sum \epsilon_{ij}R_{j}E(P_2^{2k-1}) + \sum x_j
E(S_{ij}),\cr
\Delta C(P_2^{2k})  &=& \sum x_i E(T_i).\cr
\end{eqnarray*}

The chains $Q_i, S_{ij}, T_i$ are closed and, unless $R_1$ or $R_2$ is
linear, exact. The reduction exemplified in (4.4) and in (4.5)  
is then available. The result is
$$
\Delta C(P_1^{2k}) + ...=  E(P_i^{2k-1})\partial_iR ,\quad
\Delta C(P_i^{2k+1}) + ...= E(P_2^{2k-1})\epsilon_{ij}\partial_jR,\quad
\Delta C(P_2^{2k})  = 0.
$$

(c) The last case ($P_2^{2k} \in C_{2k,k+1})$ is simpler than the others
and we give the details in that case only. Let $\tau \in S_p$ be the 
reversing permutation. Garsia's formula tells us that the chains $C_{p,k}$ 
correspond to the character $\tau \mapsto (-)^k$, so the projection
$e_{2k}(k+1)P_2^{2k}$ has $\tau \mapsto (-)^{k+1}$. 
Now $\Delta C(P_2^{2k}) = \sum_{i=1}^2x_iE(a_i)$, with
$a_i \in C_{2k-1}$ closed and with the same symmetry: $\tau \mapsto
(-)^{k+1}$. The symmetry of
$C_{2k-1,k}$ (where the homology is) is $(-)^k$; therefore $a_1$ and $a_2$
are exact. The reduction
process encounters no homology and leads to zero.
\b

Putting it all together we get the following result (for notations, 
see Propositions 2.6 and 4.8).
\b

\no{\bf 4.9. Theorem.} {\it  Let $V_R$ be the space of vector fields, with
values in $A$, that annihilate $R$. Then as vector spaces,}
\begin{eqnarray*}
H^1 &=&   V_R,\cr
H^{2k,k} &=& A/DR,\cr
H^{2k,k+1} &=& \{a\in A_+, a\partial_1R = a\partial_2R = 0\},\cr
H^{2k+1,k+1} &=&    V_R/\{AdR^*\},~~ k>0.
\end{eqnarray*}

\section{Deformation of $x^n = 0$.}

Here we complete the calculation of Hochschild cohomology
of the algebra $ A = \mathbb{C}[x]/x^n,~n\geq 2$.
This purely algebraic problem, though not associated with a curve, is
nevertheless very similar to that posed by curves. In the context of
singularity theory it is one of the standard forms. The chains are
restricted to positive degree. This subalgebra of $A$ is denoted $A_+$.
\b

\no {\it 5.1. Homology.} For convenience,   $xxx^2...$ shall stand for
either $x\otimes x\otimes x^2...$ or $x,x,x^2,...\,\, $. The spaces $H_p$
are one-dimensional for $p \geq 1$ and representative elements of $Z_p$
are $x$, $xx^{n-1}$, $xx^{n-1}x$,..., or $(xx^{n-1})^k$ for $p=2k$ and 
$(xx^{n-1})^kx$ for $p = 2k+1$.
\b

\no {\it 5.2. Closed cochains.} A $p$--cochain $C$ is closed if
$$
\delta C(a_1...a_{p+1}) := a_1C(a_-) + (-)^{p+1}a_{p+1}C(a_+) - C(da) = 0,
\eqno(5.1)
$$
with $a_- = a_2\,...\,a_{p+1},~a_+ = a_1\,...\,a_p$. We interpret this
relation, in the first place, as a recursion relation that determines the
cochain
$C$ on exact arguments, in terms of its values on arguments of lower
degree. For example, if the 1--cochain $C$ is closed, then $C(x^k) =
kx^{k-1}C(x),  ~k = 2,...,n$.
Hence $C(x^k)$ is determined for $k = 2,...,n-1$ by $C(x)$, and
$C(x)\in A_+$ (thus restricted to positive degree).

The obstruction to this interpretation of (5.1) is $da = 0$; in this case
closure requires the relation
$$
\delta C(a) = a_1C(a_-) + (-)^{p+1}a_{p+1}C(a_+) = 0.\eqno(5.2)
$$
But if $a = db$, then this last relation is automatic, since
\begin{eqnarray*}
\delta C(db) &=& b_1b_2C(b_3...) - b_1C(db_-)
+ (-)^{p+1}b_{p+2}C(db_+) - b_{p+1}b_{p+2}C(b_1...b_{p}) \cr
&=& b_1b_2C(b_3...) -b_1\biggl(b_2C(b_3...) +
(-)^{p+1}\, b_{p+2}C(b_2...b_{p+1})\biggr)\cr
&~&-  \,b_{p+1}b_{p+2}C(b_1...b_{p}) +
b_{p+2}\biggl((-)^{p+1}b_1C(b_2...b_{p+1}) +
\,b_{p+1}C(b_1...b_{p})\biggr)\cr &=& 0.
\end{eqnarray*}

The real obstruction is thus the presence of homology. When $a =
xx^{n-1}x...~$, then (5.2) reduces to
$$
p=2k:~~~~~~xC(x^{n-1}...\,x  - x...\,x^{n-1}) = 0,\eqno(5.3)
$$
$$
p=2k-1:~xC(x^{n-1}...\,x^{n-1}) + x^{n-1}C(x...\,x)=0. \eqno(5.4)
$$

\no{\it 5.3. Proposition. The obstructions to interpreting the closure
condition (5.1) as recursively fixing the value of $C(da)$ in terms of 
values of $C$ on arguments of lower degrees are:
\footnote {From now on dots indicate a sequence in which $x$ and $x^{n-1}$
alternate.}
$$
p = 2k: {\rm none}, ~~  p=2k-1: ~~x^{n-1}C(x...,x).\eqno(5.5)
$$}
Homology selects the argument here also. The truth of the Proposition is
obvious except for the possibility of accidental cancellations. Here,
nevertheless, is a direct
proof.
\b

\no{\it Proof of Proposition 5.3, case} $p = 2k$.   For $p =2k$ and 
$m = 1,2,...,\alpha,~~\alpha = k(n-2) + 1$, let
$$
\phi^m := \sum_{\matrix{ 1 \leq p_1,...\,,p_k \leq m\cr p_1 + ... + p_k =
k+m-1\cr}}
xx^{p_1}xx^{p_2}...\,xx^{p_k}x.\eqno(5.6)
$$
It may be shown by induction that
$$
d\phi^{\alpha-1} = x^{n-1}...\,x - x...\,x^{n-1},~~ d\phi^m =
\phi_-^{m+1}-\phi_+^{m+1}, ~ m<\alpha-1.
$$
Posing $\delta C(\phi^m) = 0$ for $m < \alpha$, we find that the left hand
side of (5.3)vanishes identically:
   $$
xC(x^{n-1}...\, x - x ...\,x^{n-1}) = xC(\phi_-^\alpha - \phi_+^\alpha) =
xC(d\phi^{\alpha-1}) =
x^2C(\phi_-^{\alpha-1} - \phi_+^{\alpha-1}) = ...~.
$$
Iteration ends with $x^nC(a_-^{\alpha+1-n}- a_+^{\alpha +1-n}) = 0$.
\b

\no{\it Proof of Proposition 5.3, case} $p = 2k-1$. For $m =
1,2,...\,\alpha = k(n-2) + 1$,  set
$$
\psi^m := \sum_{\matrix{ 1 \leq p_1,...\,,p_k \leq m\cr p_1 + ... + p_k =
k+m-1\cr}}
xx^{p_1}xx^{p_2}...\,xx^{p_k} .\eqno(5.7)
$$
Then
$
d\psi^{\alpha-1} = x^{n-1}...\, x^{n-1} = \psi_-^\alpha$ and for
$m<\alpha-1,~~d\psi^m =
\psi_-^{m+1}-\phi^{m+1},
$
and
$$
(x^l\psi_{p+1}^{\alpha-l})\otimes \psi_+^{\alpha-l} = x^{n-1} \otimes
\phi^{\alpha + 2 -n},~~ l =
0,1,...\,n-2.\eqno(5.8)
$$
If $\delta C(\psi^m) = 0,~~ m < \alpha$, then the left hand side of (5.4)
is
\begin{eqnarray*}
xC(x^{n-1}...\, x^{n-1}) &+& x^{n-1}C(x...\,x)\cr &=& xC(d\psi^{\alpha-1}
+ x^{n-1}C(\phi^{\alpha
+2-n})\cr & =& x^2C(\psi_-^{\alpha-1}) + 2x^{n-1}C(\phi^{\alpha +2-n}) =
... \cr & =& x^{n-1}C( \psi_-^{\alpha+2-n}) + (n-1)x^{n-1}C(\phi^{\alpha
+2-n})\cr & =& x^{n-1}C(d\psi^{\alpha + 1-n} + \phi^{\alpha + 2 n}) +
(n-1)x^{n-1}C(\phi^{\alpha + 2-n})\cr
&=& nx^{n-1}C(x...\,x).
\end{eqnarray*}

The proof of Proposition 5.3 is complete. The  implication is that, if
a $(2k-1)$--cochain $C$ is closed, then $C(x...x) \in A_+$.

\b
\no {\it 5.4. Exact cochains.} Exact $p$--cochains have the form
$$
C(a_1...a_p) = a_1E(a_-) +(-)^pa_pE(a_+) - E(da).\eqno(5.9)
$$
The obstruction to interpreting this relation as a recursion relation to
determine the $E(da)$ is $da = 0$. Here too, the real
obstruction, when $C$ is closed, is the existence of homology.
The most useful interpretation is this. Given any closed $p$--cochain
a ``gauge transformation" is the addition of an exact $p$--cochain,
$C\rightarrow C + \Delta C$, with
$$
\Delta C(a_1...a_p) = a_1E(a_-) +(-)^pa_pE(a_+) - E(da).\eqno(5.10)
$$
The space $H^p$ is the space of gauge invariant evaluations of closed
$p$--cochains.

To illustrate, here is the situation for 2--cochains, when $n = 3$. Closure,
$$
\delta C(xxx) = C(xx^2) - C(x^2x) = 0,~~~\delta C(xx^2x) = xC(x^2x-xx^2)= 0.
$$
Gauge transformation
$$
\Delta C(xx) = 2xE(x)-E(x^2),~~~\Delta C(xx^2) = xE(x^2) + x^2E(x) =
\Delta C(x^2x),
$$
By means of gauge transformations we can, for example, reduce $C(xx)$ to zero.
Cohomology is the existence of the gauge invariant object
$C(xx^2)+xC(xx)$ Mod~$x^2$.
\b

\no{\it 5.5. Theorem.  The space of the gauge-equivalent evaluations, and
the associated cohomology spaces
on $Z_p(A,A)$ are as follows}
\begin{eqnarray*}
p = 0&:& A\cr
& &H^0(A,A) = {\rm span}\{1,x,...,x^{n-1}\}, ~{\rm dim.} =1;\cr
p = 1&:&C(x)\cr
    & &H^1(A,A) = {\rm span}\{x,...,x^{n-1}\},~{\rm dim.} =n-1;\cr
p=2k-1&:&\sum_{l=0}^{n-2}x^lC(\phi^{\alpha-l})~~~(k>1)\cr
   & &H^{2k-1}(A,A) = {\rm span}\{x,...,x^{n-1}\},~{\rm dim.} =
n-1;\cr
p=2k&:& \sum_{l=0}^{n-2}x^lC(\psi^{\alpha-l})~~{\rm Mod} ~\mathbb{C} x^{n-1}\cr
& &H^{2k}(A,A) = {\rm span}\{1,x,...,x^{n-2}\},~ {\rm dim.} =
n-1.\cr
\end{eqnarray*}
\b

\no{\it Proof.} By a direct and straightforward calculation we obtain, 
for $p = 2k$,\linebreak
$
\sum_{l=0}^{n-2}x^l\Delta C(\psi^{\alpha-l}) = nx^{n-1}E(x...\,x),
$
and for $p= 2k-1$, ~$\sum_{l=0}^2 x^l\Delta C(\phi^{\alpha-1}) = 0$.
\b

\no{\it 5.6. Proposition. The BGS ~${}^`$\hskip-.1mmdecomposition' ~for
$k\geq 1$ is}
$$
H_{2k} = H_{2k,k},~~~ H_{2k+1} = H_{2k+1,k+1}.
$$
\b
\no{\it Proof.} The element $x...x^{n-1}\in Z_{2k}$ lifted to
$Z_{2k}(\mathbb{C}[x],\mathbb{C}[x])$, is
$$
d(xx^{n-1})^k = x^n \bigcirc \hskip-4mm sh (xx^{n-1})^{k-1}.
$$
If $(xx^{n-1})^{k-1}$ is of type $H_{2k-2,k-1}$, then the right hand side
is of type $H_{2k-1,k}$,
and $(xx^{n-1})^k$ is of type $H_{2k,k}$. Since $xx^{n-1}$ is indeed of
type $H_{2,1}$ the result
follows by induction. Similarly,
$
d(xx^{n-1})^kx = x^n \bigcirc \hskip-4mm sh (xx^{n-1})^{k-1}x,
$
and the same argument applies {\it mutatis mutandi}.


\renewcommand{\a}{{\alpha}}

\b\b\b
\ce{\stepthree Appendix}
\b 
   \ce{\bf   Hochschild and Harrison cohomology of complete
intersections}
\b

I will explain here a way to calculate Hochschild and
Harrison cohomology groups for algebras of functions on singular planar 
curves etc. based on Koszul resolutions.
This calculation is standard and definitely known to specialists.

\b
\no{\bf A1. Reminder on complete intersections and Koszul resolution}
\b

Results of this section can be found e.g. in the classical textbook \cite{Ma}.

Suppose that we are given a system of polynomial equations
(say, over the field of complex numbers $\C$, one can replace it by an
arbitrary field):

$$f_1(z_1,\dots,z_n)=0,\dots,f_m(z_1,\dots,z_n)=0$$

Denote by $A$ the quotient algebra $P/(f_1,\dots,f_m)$ where $P$
denotes the ring of polynomials
   $\C[z_1,\dots,z_n]$.

We say that we have a complete intersection if
the dimension of the set of solutions of the system above is $n-m$.
A sufficient condition for this is that
   $f_1,\dots,f_m$ form a regular sequence in $P$,
   i.e. for any $k\le n$ element $f_k$ is not a divisor of zero
in the quotient of $P$ by the ideal generated by $f_1,\dots,f_{k-1}$.

\def\A{\tilde A}

\begin{thm} Assume (in the previous notations)
    the condition of the complete intersection.
   Let us consider $\Z_{\le 0}$--graded supercommutative
superalgebra $$\A:=P\otimes\wedge(\{\a_j\}_{j=1,\dots,m})$$
where subalgebra $P$ is in degree $0$ and generators $\a_j$ are
in degree $-1$, endowed with differential
$$
d_{\A}  :=\sum_j f_j \frac{\partial}{\partial \a_j}.
$$  Then cohomology of
this differential
is zero in negative degrees and isomorphic to $P/(f_1,\dots,f_m)$ in
degree $0$.
\end{thm}

   In the above theorem one can replace $P=\C[z_1,\dots,z_n]$ by the algebra
   of functions on arbitrary smooth $n$--dimensonal affine algebraic variety.
Complex $(\A,d_{\A})$ is called the Koszul resolution of $A$.
    Slightly abusing notations we will write
      $\A=\C[z_1,\dots,z_n;\a_1,\dots,\a_m]$ meaning that $(\a_i)$ are
fermionic (odd) variables.
   Here and later variables denoted by Latin (resp. Greek) letters are
even (resp. odd).

\b\b

\no{\bf A2. Generalities on Hochschild and Harrison cohomological complexes
for differential graded algebras}

In what follows all complexes will be $\Z$--graded with the
differential of degree $+1$.
A morphism of complexes is called a quasi-isomorphism iff it induces
an isomorphism of cohomology groups. A vector space
can be considered as a complex concentrated in degree $0$ and
endowed with zero differential.

Definitions of homological and  cohomological Hochschild complexes extend 
immediately to the case of differential graded algebras (dga in short), the 
same for Harrison (co-)homological complexes in the graded commutative case.
The underlying $\Z$--graded space for the
cohomological Hochschild complex for a dga $F$ with coeffcients in 
a dg bimodule $M$ is defined as the infinite product (in the category of 
$\Z$--graded spaces)
     $$C^\bullet(F,M):=\prod_{n\geq 0} \underline{Hom}(F[1]^{\otimes n},M)$$
where $\underline{Hom}$ is inner Hom--space in tensor category of
$\Z$--graded spaces,
$$\left(\underline{Hom}(U,V)\right)^k:=\prod_{n\in\Z} Hom(U^n,V^{n+k})$$
and $F[1]$ denotes the complex obtained from $F$ by the shift of the grading, 
$F[1]^k:=F^{k+1}$. The formula for the differential in $C^\bullet(F,M)$ 
is the sum of a super-version of the formula for the differential in the  
an ordinary algebra (in degree 0), and a term arising from the
differential in $F$ itself (see e.g. section 5.3 from \cite{L} for a similar 
case of the homological Hochschild complex).

\begin{lmm} \label{lmm:coh}If $\phi:\F\ra F$ is a quasi-isomorphism
between two dga's, then the corresponding cohomological Hochschild complexes 
$C^{\bullet}(F,F)$ and $C^{\bullet}(\F,\F)$ are quasi-isomorphic.
\end{lmm}
   \emph{Proof:}
An algebra $F$ can be considered as a differential graded bimodule
over $\F$ via the homomorphism $\phi:\F\ra F$.
Let us consider three complexes and natural homomorphisms between them:
    $$C^\bullet (\F,\F)\ra C^\bullet(\F,F)\la C^\bullet(F,F)$$
 All three complexes carry complete decreasing filtrations
with the associated quotients (and maps between them)
      $${\underline{Hom}}(\F[1]^{\otimes k},\F)\ra
{\underline{Hom}}(\F[1]^{\otimes k},F)\la
      {\underline{Hom}}(F[1]^{\otimes k},F)$$
We see that associated quotients are quasi-isomorphic, and
applying spectral sequences we conclude that $C^\bullet (\F,\F)$ and 
$C^\bullet(F,F)$ are quasi-isomorphic.\qed

For a graded supercommutative $F$ one can define the Hodge
decomposition for Hochschild cochains, and Harrison cohomology
in the same way as in the usual non-graded case. In the above lemma 
the quasi-isomorphism between Hochschild cohomology of the 
resolution and of algebra itself is manifestly compatible
with the Hodge decomposition.
   \b

\no{\bf A3. Calculation of Hochschild and Harrison cohomology for complete
intersections}

The cohomological Hochschild--Kostant--Rosenberg theorem says that
the Hochschild cohomology of the algebra $\mathcal{O}_X$ of functions 
on an algebraic affine manifold $X$ is the algebra $T_X^{poly}$ of 
polyvector fields on $X$. Moreover, there is a canonical quasi-isomorphism 
$T_X^{poly}\ra C^\bullet(\mathcal{O}_X,\mathcal{O}_X)$
mapping polyvector field $f \,v_0\wedge\dots\wedge v_n$ where
$f\in \mathcal{O}_X$, $(v_i)_{i=1,n}$ are derivations of $\mathcal{O}_X$, 
to the polylinear operator
$$a_1\otimes\dots \otimes a_n\mapsto f\sum_{\sigma\in \Sigma_n}sign(\sigma)
\prod_i v_{\sigma(i)}(a_i)$$

The super-version of this theorem is also true, e.g. for supermanifold 
$Y=\C^{n|m}$, we have $\mathcal{O}_Y=\C[z_1,\dots,z_n;\a_1,\dots,\a_m]$ 
and its Hochschild cohomology  is the algebra $T_{Y}^{poly}$:
$$T_Y^{poly}=\C[z_1,\dots, z_n;\eta_1,\dots,\eta_n;\a_1,\dots,\a_m;b_1,\dots,b_m],$$
$$deg(z_i)=0\,\,\,deg(\eta_i)=+1,\,\,\,deg(\a_j)=-1,\,\,\,deg(b_j)=+2$$
Here the new variables $\eta_i,\,b_j$ have the meaning of derivations
$\partial/\partial z_i,\,\partial/\partial \a_j$.
Strictly speaking, here we should consider not polynomials  but formal 
power series with respect to  variables $\eta_i, b_j$, but it gives the same 
result in the category of $\Z$-graded spaces because there are only finitely 
many monomials in $\eta_i, b_j$ in any given degree.

The dga $\A$ is obtained from $\mathcal{O}_Y$ by ``switching on'' 
the differential $d_{\A}$. Here we describe  the corresponding HKR 
description of the Hochschild cohomology of $\A$, and therefore of 
  $H^\bullet(A,A)$ by lemma 1.

\begin{prop} Complex $C^\bullet(\A,\A)$ is quasi-isomorphic to
$T:=T_Y^{poly}$ endowed with the differential
         $$ d_T:=\sum_{i,j}{\frac{\partial f_j}{\partial z_i}} b_j
{\frac{\partial}{\partial  \eta_i}} +
\sum_j f_j{\frac{\partial}{\partial \a_j}}$$
The Hodge grading is given by counting variables $\eta_i, b_j$.\end{prop}

\emph{Proof:}      The formula for $d_T$ is just the formula for
   the Lie derivative of a polyvector field on $Y=\C^{n|m}$ with respect 
to the odd vector field $d_{\A}=\sum_j f_j\frac{\partial}{\partial \a_j}$.
It is easy to see that the formulas from above give a homomorphism of complexes
       $$\chi:(T, d_T)\to C^\bullet(\A,\A)$$
   We have to prove that it is a quasi-isomorphism.   
   Let us introduce $\Z_{ \geq 0}$--grading $deg_\a$ on  $\mathcal{O}_Y$ by 
the total number of variables $\a_j$ (incidentally, it coincides
with minus the standard $\Z$--grading on $\mathcal{O}_Y$). A Hochschild cochain
$\mathcal{O}_Y^{\otimes n}\ra \mathcal{O}_Y$ is called homogeneous of
    $\deg_\a$ degree $N\in \Z$ if it is homogenous with respect to
grading $deg_\a$ of degree $N$.
    The whole Hochschild complex $C^\bullet(\mathcal{O}_Y,\mathcal{O}_Y)$ 
is the product over all $N\in \Z$ of subcomplexes consisting of $\deg_\a$ 
degree $N$ cochains.
  The Hochschild differential of algebra $\mathcal{O}_Y$ preserves the 
$\deg_{\a}$ grading. The correction to the differential coming from 
$d_{\A}$ decreases this grading by $1$. Finally, it is obvious that for a 
non-zero cochain its $\deg_\a$ is bounded from below (by $-m$).
Therefore we have a convergent spectral sequence proving that $\chi$ is a
quasi-isomorphism. The statement about the Hodge grading is obvious. \qed

Now we introduce a smaller complex
$$\tilde{T}:= A[\eta_1,\dots, \eta_n;b_1,\dots,b_m],\,\,\,
d_{\tilde{T}}:=\sum_{i,j}{\frac{\partial f_j}{\partial x_i}} b_j
{\frac{\partial}{\partial  \eta_i}}$$
where the variables have the same grading as before, 
$deg(\eta_i)=+1,\,deg(b_j)=+2$.

\begin{thm} Under the previous assumptions the Hochschild cohomology
of $A$ is isomorphic to the cohomology
   of complex $(\tilde{T},d_{\tilde{T}})$. The Hodge grading is given
by counting variables $\eta_i, b_j$.
\end{thm}

\emph{Proof:} The obvious map  $(T_F,d_T)\ra (\tilde{T},d_{\tilde{T}})$ 
induces a quasi-isomorphism on graded quotients for the filtration 
by the total number of variables $\eta_i$. \qed

The conclusion for the only non-trivial Harrison cohomology are in
degrees $1$ and $2$ and are given by kernel and cokernel of the map
   $$A^n \stackrel{\left(\partial f_j/\partial_{z_i}
\right)}{\longrightarrow} A^m$$
In particular, there is no obstruction for commutative deformations
as $Harr^3(A)=0$.
It is easy to see that a miniversal commutative deformation of $A$ is
given by any deformation $\tilde{f}_1(z,t),\dots,\tilde{f}_m(z,t)$
of polynomials $f_1(z),\dots, f_m(z)$ depending on formal parameters
$t_1,\dots,t_N$ where $N=rk\,\,Harr^2(A)$, such that vectors
$$v_k:=\left({\frac{\partial \tilde{f}_1}{\partial t_k}}_{|t=0},\dots,
{\frac{\partial \tilde{f}_m}{\partial t_k}}_{|t=0}\right),
\,\,\,k=1,\dots,N$$
form a basis in $Harr^2(A)=A^m/\left(\frac{\partial f_j}{\partial
z_i}\right) A^n$.
The deformed algebra is
$$\C[[t_1,\dots,t_N]][z_1,\dots,z_n]/I$$ 
where $I$ is the completion with respect to 
the  topology on $\C[[t_1,\dots,t_N]]$ associated with the maximal ideal,
of the ideal generated by $\tilde{f}_1(z,t),\dots,\tilde{f}_m(z,t)$.

In particular, if we have only one equation $f(z)=f_1(z)=0$
then $Harr^2(A)$ is the quotient 
$\C[z_1,\dots,z_n]/\left(f,\partial f/\partial
z_1,\dots, \partial f/\partial z_n\right)$.

In the case $n=2$ and $m=1$, Hochschild cohomology groups consists of
an unstable part in lower degrees and $2$--periodically repeated block
$$A\stackrel{(\partial_{z_1} f_1,\partial_{z_2}
f_1)}{\longrightarrow} A\oplus A
\stackrel{(\partial_{z_2} f_1,-\partial_{z_1} f_1)}{\longrightarrow} A$$

Finally, for $n=m=1$, $A=\C[z]/(z^k)$ we have
$$H^O(A,A)=A\simeq \C^k,\,\,\,H^l(A,A)\simeq \C^{k-1} \mbox{ for }l=1,2,\dots$$

\no{\bf A4. Calculation of Hochschild a homology with coefficients with the
diagonal bimodule, for complete intersections}

   Similarly, one can calculate Hochschild  homology $H_*(A,A)$
   for complete intersections. Here is the final result:

\begin{thm} In previous notations and under the  assumption of
   complete intersection the Hochschild homology $H_*(A,A)$ of $A$ is
isomorphic to the cohomology of complex
$\tilde{\Omega}:=A[\xi_1,\dots,\xi_n;a_1,\dots,a_m]$ where degrees of 
variables are $deg(\xi_i)=-1,\,\, \,deg(a_j)=-2$ endowed with the differential
   $d_{\tilde{\Omega}}:=\sum_{i,j}{\frac{\partial f_j}{\partial z_i}}
\xi_i {\frac{\partial}{\partial  a_j}}$.
    The Hodge grading is given by counting variables $\xi_i, a_j$.
\end{thm}
   The proof is parallel to one for the cohomological complex.
   An example of this calculation for the case of truncated polynomial ring
   can be found in \cite{L}, exercise E.4.1.8 and Proposition 5.4.15.


\begin{thebibliography}{9999}


\bibitem{B} 
Barr M., Harrison homology, Hochschild homology and triples, 
J. Algebra {\bf 8} (1968) 314--323.\\
\textit{Cohomology of commutative algebras}, Doctoral dissertation, 
U.Penn. 1962.



\bibitem{BFFLS} 
Bayen F., Fronsdal C., Flato M., Lichnerowicz A. and Sternheimer D.,
Quantum Mechanics as a Deformation of  Classical Mechanics,
Annals of Physics {\bf 111} (1978 ) 61--110 and 111--151.


\bibitem{DFST} 
Dito G., Flato M., Sternheimer S. and Takhtajan L., 
Deformation Quantization and Nambu Mechanics, Comm. Math. Phys.
{\bf 183} (1997) 1--22.


\bibitem{Fe} 
Fedosov B.V.,  A simple geometrical construction of
deformation quantization,
J. Diff. Geom. {\bf 40} (1994) 213--238, and earlier papers.


\bibitem{Fl}
Fleury  P.J., Splittings of Hochschild's complex for commutative
algebras, Proc. Amer. Math. Soc. {\bf 30} (1971) 405--323.


\bibitem{Fr} 
Fronsdal C., Some ideas about quantization, Rep. Math. Phys.
{\bf 15} (1978) 111--145.

\bibitem{Fr2}
Fronsdal C., Harrison cohomology and abelian deformation quantization on 
algebraic varieties. \textit{Deformation quantization} (Strasbourg, 2001), 
149--161, IRMA Lect. Math. Theor. Phys. {\bf1}, de Gruyter, Berlin 2002.\\
Abelian Deformations, Proceedings of the IX$^\mathrm{th}$ International 
Conference on Symmetry Methods in Physics, Yerevan, July 2001.

\bibitem{Ga}
Garsia A.M., Combinatorics of the free Lie algebra and the
symmetric group, \textit{Analysis, et cetera}, Research Papers 
Published in Honor of J\"urgen Moser's 60$^\mathrm{th}$ Birthday, 309--392, 
Academic Press, New York 1990.


\bibitem{G1}
Gerstenhaber M., On the deformations of rings and algebras,
Annals of Math. {\bf 79} (1964) 59--103.


\bibitem{G2}
Gerstenhaber M., Developments from Barr's thesis, presented at
the celebration of M. Barr's 60$^\mathrm{th}$ birthday, June 30, 1998.


\bibitem{GS}
Gerstenhaber M. and Schack S.D., A Hodge-type decomposition for
commutative algebra cohomology, J. Pure Appl. Algebra
{\bf 48} (1987) 229--247.


\bibitem{H} 
Harrison D.K., Commutative algebras and cohomology,
Trans. Amer. Math. Soc. {\bf 104} (1962) 191--204.


\bibitem{HKR}
Hochschild G., , Kostant B. and Rosenberg A., Differential
forms on regular affine algebras, Trans. Amer. Math. Soc. {\bf 102}
(1962) 383--408.


\bibitem{Kon}
Kontsevich M., Deformation quantization of Poisson
manifolds, Lett. Math. Phys. \textbf{66} (2003), 157--216 
(\texttt{q-alg/9709040}).\\
\bibitem{Kon2}
Kontsevich M., Operads and motives in deformation quantization, 
Lett. Math. Phys. {\bf 48} (1999) 35--72.

\bibitem{Kos} 
Kostant B., {\it  Quantization and unitary representations,
Lectures in Modern Analysis and Applications III}, Lecture Notes in
Mathematics {\bf 170},   87--208, Springer-Verlag, Berlin, 1970.


\bibitem{L} 
Loday  J.-L. {\em Cyclic homology}, 
Grundlehren der Mathematischen Wissenschaften {\bf 301}, xx+513 pp,
Springer-Verlag, Berlin 1998.


\bibitem{Ma}
Matsumura, H. {\em Commutative algebra}, W.A. Benjamin Co., New
York, 1970.


\bibitem{M}
Moyal J.E., Quantum mechanics as a statistical theory,
Proc.Cambridge Phil. Soc. {\bf 45} (1949) 99--124.


\bibitem{OMY}
Omori H., Maeda Y. and Yoshioka A., Existence of closed star
products, Lett. Math. Phys. {\bf 26} (1992) 285--294.


\bibitem{S}
Souriau J.M., {\it Structures des Syst\`emes Dynamiques},
Dunod, Paris, 1970.


\bibitem{T} 
Tamarkin D.E., Formality of chain operad of little discs,  
Lett. Math. Phys. {\bf 66}  (2003),  65--72. See also:
Another proof of M. Kontsevich's formality
theorem for $\mathbb{R}^n$, \texttt{math.QA/9803025}.


\bibitem{V}
Vey J.,  D\'eformation du crochet de Poisson sur une vari\'et\'e
symplectique, Comment. Math. Helv.  {\bf 50} (1975) 421--454.


\bibitem{W}
Weyl H., {\it  Theory of Groups and Quantum Mechanics},
Dover, New York, 1931.
\end{thebibliography}
\end{document}